\documentclass[preprint,authoryear,12pt]{elsarticle}

\usepackage{graphicx}

\usepackage{amssymb}

\usepackage[nodots]{numcompress}

\journal{New Astronomy}

\newcommand{\Mwd}{\mbox{$M_\mathrm{wd}$}}
\newcommand{\Msec}{\mbox{$M_\mathrm{sec}$}}
\newcommand{\Porb}{\mbox{$P_{\rm orb}$}}
\newcommand{\Psh}{\mbox{$P_{\rm sh}$}}

\begin{document}

\begin{frontmatter}



\title{The orbital and superhump periods of the dwarf nova\\
HS\,0417+7445 in Camelopardalis}


\author[1]{J.H. Shears}
\author[2]{B.T. G\"ansicke}
\author[3]{S. Brady}
\author[4]{P. Dubovsk\'y}
\author[5]{I. Miller}
\author[6]{B. Staels}

\address[1]{BAA, "Pemberton", School Lane, Bunbury, Tarporley, Cheshire, CW6 9NR, UK}
\address[2]{Department of Physics, University of Warwick, Coventry, CV4 7AL, UK}
\address[3]{AAVSO, 5 Melba Drive, Hudson, NH 03051, USA}
\address[4]{Astronomical Observatory on Kolonica Saddle, Slovakia}
\address[5]{BAA, Furzehill House, Ilston, Swansea, SA2 7LE, UK}
\address[6]{CBA Flanders, Alan Guth Observatory, Koningshofbaan 51, Hofstade, Aalst, Belgium}

\begin{abstract}
We present the 2005--2010 outburst history of the SU\,UMa-type dwarf
HS\,0417+7445, along with a detailed analysis of extensive time-series
photometry obtained in March 2008 during the second recorded
superoutburst of the system. The mean outburst interval is
$197\pm59$\,d, with a median of 193\,d. The March 2008 superoutburst
was preceeded by a precursor outburst, had an amplitude of 4.2
magnitudes, and the whole event lasted about 16 days. No superhumps
were detected during the decline from the precursor outburst, and our
data suggests instead that orbital humps were present during that
phase. Early superhumps detected during the rise to the superoutburst
maximum exhibited an unusually large fractional period excess of
$\epsilon=0.137$ ($\Psh=0.0856(88)$\,d). Following the maximum, a
linear decline in brightness followed, lasting at least 6 days. During
this decline, a stable superhump period of $\Psh=0.07824(2)$\,d was
measured. Superimposed on the superhumps were orbital humps, which
allowed us to accurately measure the orbital period of HS\,0417+7445,
$\Porb=0.07531(8)$\,d, which was previously only poorly estimated. The
fractional superhump period excess during the main phase of the
outburst was $\epsilon=0.037$, which is typical for SU\,UMa dwarf
novae with similar orbital period. Our observations are consistent
with the predictions of the thermal-tidal instability model for the
onset of superoutbursts, but a larger number of superoutbursts with
extensive time-series photometry during the early phases of the
outburst would be needed to reach a definite conclusion on the cause
of superoutbursts.
\end{abstract}


\end{frontmatter}


\section{Introduction}
In cataclysmic variables (CVs) a white dwarf primary accretes material
from a secondary star via Roche lobe overflow. The secondary is
usually a late-type main-sequence star \citep{warner95-1}.  In the
absence of a significant white dwarf magnetic field, material from the
secondary is processed through an accretion disc before settling on
the surface of the white dwarf. In CVs with low to intermediate mass
transfer rates, dwarf nova outbursts with amplitudes of $2-8$\,mag and
durations of days to weeks are observed. The outbursts are thought to
be caused by a thermal instability in the accretion disc associated
with partial ionisation of hydrogen \citep{meyer+meyer-hofmeister81-1,
  smak81-2, cannizzoetal82-1}, and such time-dependent models for accretion
disc evolution have been applied to a variety of astrophysical
environments, e.g. AGN \citep{burderietal98-1} or young stellar
objects \citep{bell+lin94-1}. Dwarf novae of the SU\,UMa family
occasionally exhibit superoutbursts which last several times longer
than normal outbursts and may be up to a magnitude brighter. During a
superoutburst the light curve of a SU\,UMa system is characterised by
superhumps. These are modulations in the light curve which are a few
percent longer than the orbital period. They are thought to arise from
the interaction of the secondary star orbit with a slowly precessing
eccentric accretion disc. The eccentricity of the disc arises because
a 3:1 resonance occurs between the secondary star orbit and the motion
of matter in the outer accretion disc. The actual trigger of the
superoutbursts is still debated \citep{osaki+meyer03-1,
  schreiberetal04-1}.

HS\,0417+7445 (hereafter HS\,0417), also known as
1RXS\,J042332+745300, was independently identified as a CV by
\citet{wuetal01-2} in the course of follow-up spectroscopy of objects
from ROSAT Bright Source Catalogue \citep{vogesetal99-1}, and by
\citet{aungwerojwitetal06-1} because of the presence of strong Balmer
emission lines in its optical spectrum in the Hamburg Quasar Survey
(HQS, \citealt{hagenetal95-1}).  HS\,0417 showed large amplitude
variability on the HQS spectroscopic plates, between $B\simeq18.0$ and
$B\simeq13.7$, suggesting a dwarf nova classification of the
object. Further observations of HS\,0417 by
\citet{aungwerojwitetal06-1} between December 2000 and January 2005
showed the object near to a mean magnitude of $\sim17.5$, except
during January 2001 when the system was found in outburst near
$B\simeq13.5$. Photometry during this outburst revealed superhumps
that identify HS\,0417 as a SU\,UMa type dwarf nova. Two possible
superhump periods were identified, $\Psh=108.3$ or 111.2\,min (0.0752
or 0.0772\,d). Analysis of the quiescent data revealed a double-humped
light curve with two possible values of the orbital period,
$\Porb\simeq 105.1$ or 109.9 min (0.0723 or 0.0763\,d). Brief
descriptions of two superoutbursts in 2008 and 2010 were given by
\citet{katoetal09-1,katoetal10-1}.

Here we report a detailed analysis of time-series photometry obtained
in 2008 during the second superoutburst of HS\,0417, from which we
determine accurate orbital and superhump periods. We also discuss the
possible mechanisms triggering superoutbursts, and how pre-cursor
outbursts, such as detected in our data, can help to distinguis
between the competing models. 

\begin{table}[t]
\caption[]{Log of the observations\label{t-obslog}.}
\setlength{\tabcolsep}{0.95ex}
\begin{flushleft}
\begin{tabular}{lcccc}
\hline\noalign{\smallskip}
Date      &   Start time    &  Duration & Filter & Observer\\
2007 (UT) &   (JD--2454000) &  (h)      &        &         \\
\hline\noalign{\smallskip}
Mar 4   & 530.304 & 4.8  & C &   JS \\
Mar 4   & 530.363 & 5.3  & V &   IM \\
Mar 5   & 531.337 & 0.5  & C &   BS \\
Mar 6   & 531.558 & 7.0  & C &   SB \\
Mar 6   & 532.247 & 10.0 & V &   PD \\
Mar 7   & 532.546 & 7.5  & C &   SB \\
Mar 7   & 533.237 & 5.8  & V &   PD \\
Mar 7   & 533.302 & 4.1  & V &   IM \\
Mar 7   & 533.398 & 3.2  & C &   JS \\
Mar 7   & 533.441 & 4.6  & C &   BS \\
Mar 8   & 533.707 & 1.0  & V &   IM \\
Mar 8   & 534.295 & 4.4  & C &   JS \\
Mar 8   & 534.370 & 5.0  & C &   IM \\
Mar 9   & 535.292 & 2.2  & C &   JS \\
Mar 9   & 535.304 & 0.4  & C &   IM \\
Mar 9   & 535.337 & 6.7  & C &   BS \\
Mar 10  & 536.369 & 2.1  & C &   JS \\
Mar 11  & 536.506 & 7.9  & C &   SB \\
Mar 12  & 538.463 & 1.7  & C &   IM \\
Mar 13  & 538.511 & 6.6  & C &   SB \\
Mar 16  & 542.319 & 3.3  & C &   IM \\
Mar 17  & 543.325 & 4.0  & C &   IM \\
Mar 18  & 543.512 & 7.3  & C &   SB \\
\noalign{\smallskip}\hline
\end{tabular}
\end{flushleft}
Notes on the equipment used for the observations. JS: 0.28\,m SCT plus
Starlight Xpress SXV-M7; SB: 0.4\,m reflector plus SBIG ST-8XME; PD:
0.265\,m reflector plus Meade DSI Pro; IM: 0.35\,m SCT plus Starlight
Xpress SXVF-H16; BS: 0.28\,SCT plus Starlight Xpress MX716.
\end{table}

\section{Observations}

We detected HS\,0417 in outburst on 2008 March 3.977 at a filterless
magnitude of 14.5C. Time resolved photometry was conducted during the
course of the outburst according to the observation log shown in
Table\,\ref{t-obslog}. Raw images were dark-subtracted and flat
fielded before being measured using differential aperture photometry
relative to comparison stars with $V$-band photometry given by
\citet{henden10-1}. We will generally refer to dates in the truncated
form JD\,=\,JD\,--\,2454000. From the overall light curve of the
superoutburst  shown in Fig.\,\ref{f-outburst} it is apparent that
the superoutburst was preceeded by a precursor outburst. Detailed
views of the individual light curves are shown in
Fig.\,\ref{f-lightcurves}. 

\section{Analysis \& Results}
\subsection{Outburst history}

Examination of the AAVSO International Database \citep{henden10-1},
supplemented with data from the authors, reveals at least 11 outbursts
of HS\,0417 between April 2005 and October 2010
(Table\,\ref{t-outbursts}).  Two of the outbursts are definitely
superoutbursts as superhumps were recorded: the one in March 2008
outburst discussed in this paper and the 2010 Sep discussed by
\citet{katoetal10-1}. In both cases the outburst lasted about 16 days.

The April 2005 outburst lasted $\sim$8 days and the star showed a
rapid decline with no superhumps confirming this to be a normal
outburst. Similarly, the absence of superhumps during the October 2007
outbursts suggests it too was a normal outburst. Whilst we are not
aware of any time series photometry conducted during the other
outbursts, their short outburst duration (less than 6 to 8 days)
suggests they too were normal outburst.

Dwarf novae are known to exhibit quasi-period outbursts. In the case
of HS 0417, the mean outburst interval between the 11 outbursts was
$197\pm59$\,d and the median interval was 193\,d. Although
observational coverage of HS\,0417 has been good since the 2005
outburst, there are nevertheless 43 gaps in the data of more than 6 d
during which further outbursts might have been missed.  We note that
the normal outbursts of HS\,0417 have a rather long duration compared
to other SU\,UMa dwarf novae. Close monitoring of this object should
continue to improve the statistics of the outburst frequency and
duration.

\begin{figure}
\centerline{\includegraphics[angle=-90,width=0.6\columnwidth]{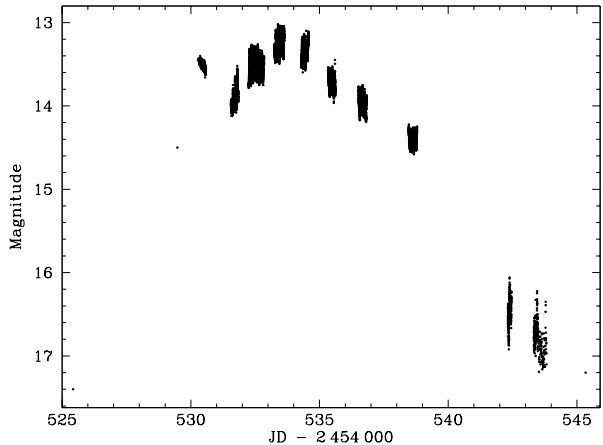}}
\caption{\label{f-outburst} Light curve of HS\,0417 during the 2008
  March outburst.}
\end{figure}

\subsection{The March 2008 superoutburst}
Figure\,\ref{f-outburst} shows the overall light curve of the
outburst, from its discovery at 14.5C on JD\,529. Photometry conducted
the next night (JD\,530) showed that the star had brightened to
$\sim13.4$C, but that it was fading at an average rate of
$0.50\,\mathrm{mag\,d^{-1}}$. By contrast, the following night
(JD\,531) the star, although considerably fainter at $\sim$14.1C, was
re-brightening rapidly at $0.91\,\mathrm{mag\,d^{-1}}$. Superhumps
were plainly visible at this point, suggesting this to be the start of
the superoutburst. It appears that the activity on JD\,529 to 530
represented a normal outburst which was the precursor, perhaps even
the trigger, for the superoutburst (see Sect.\,\ref{s-discussion}). At
its brightest on JD\,533 the star reached magnitude 13.2 (averaged
over the superhumps). An approximately linear decline followed between
maximum on JD\,533 and JD\,538 at $0.14\,\mathrm{mag\,d^{-1}}$, which
is typical of a dwarf nova in decline from a superoutburst. There was
then a 4\,d gap in the record until JD\,542 when the star was at
16.5C; hence it is not possible to conclude whether there was a
gradual or a sudden fade in the intervening period.  Finally the star
was found at magnitude 17.2C, close to its quiescence magnitude of
17.4C, on JD\,545 approximately 16 days after the outburst was first
detected. Thus the overall outburst amplitude was 4.2
magnitudes.

\begin{figure*}
\includegraphics[width=\columnwidth]{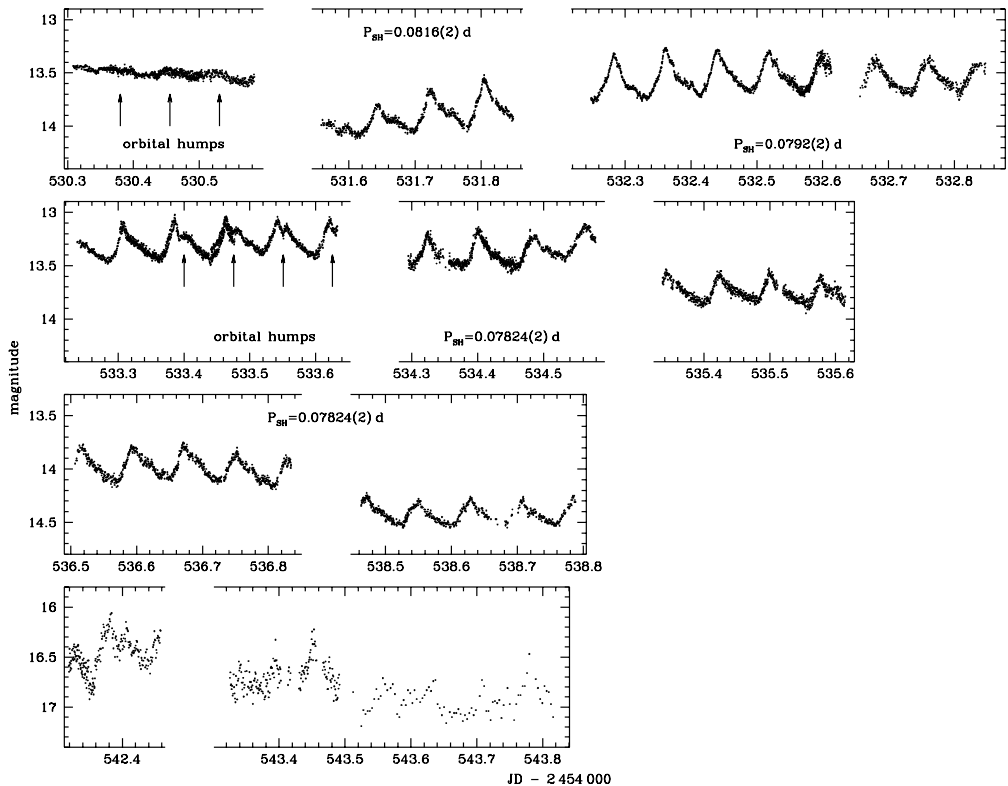}
\caption{\label{f-lightcurves} Detailed light curves of HS\,0417
  during the 2008 March outburst. All panels have the same scale in
  the time and magnitude axes to facilitate like-for-like comparison
  of the superhumps and orbital humps. Arrows indicate the occurence
  of orbital humps. The evolution of the superhump period throughout
  the outburst is indicated.}
\end{figure*}

\subsection{Superhump periods}
\label{s-psh}
The light curves obtained during the 2008 March superoutburst of
HS\,0417 are shown in detail in Fig.\,\ref{f-lightcurves}, where we
used the same scale for both time and magnitude for ease of
comparison.  Modulations in the light curve were detected throughout
the outburst, however, as we shall see, their period appeared to be
different during different stages of the outburst. Hence we analysed
different phases of the outburst light curve in turn.

Superhumps were first detected during the rise to superoutburst on
JD\,531, when they were increasing in peak-to-peak amplitude from 0.3
to 0.4\,mag (Fig.\,\ref{f-lightcurves}). 

We searched for periodic signals in our time resolved photometry using
Scargle's (\citeyear{scargle82-1}) method as optimised by
\citet{horne+baliunas86-1}, and found $\Psh=0.0856(88)$\,d. 
The following day, JD\,532, the superhumps were of the same amplitude,
but the superhump period was slightly shorter at 0.0792(22)\,d.

During the interval covering the outburst maximum on JD\,533 and the
linear decline to JD\,538 the superhump amplitude gradually diminished
from 0.4 to 0.3 magnitudes. In order to study the superhump behaviour
during this period, we first extracted the times of each resolvable
superhump maximum from the individual light curves according to the
method of \citet{kwee+vanwoerden56-1}. Times of 26 superhump maxima
were found and these were then used to assign superhump cycle numbers
which best fitted the assumption of a constant superhump period. We
found that the maxima fitted well a constant superhump period $\Psh =
0.07824(2)$\,d (112.67\,min) having the following ephemeris:

\begin{equation}
\mathrm{JD(max)} =  2454533.30893 + 0.07824(2) \times E 
\end{equation}

The superhump cycle number, the measured times of superhump maximum
and the O--C (Observed-Calculated) residuals relative to the above
superhump maximum ephemeris are shown in Fig.\,\ref{f-oc}.

\begin{figure}
\centerline{\includegraphics[angle=-90,width=0.6\columnwidth]{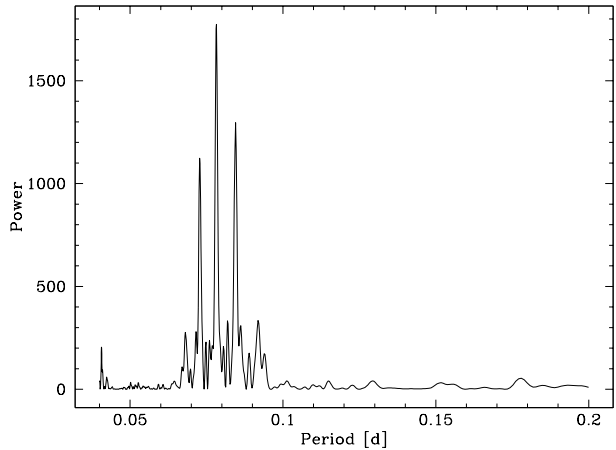}}
\caption{\label{f-power} Power spectrum of all data from JD\,533 to 538.}
\end{figure}

As an independent approach for the measurement of $\Psh$, we carried
out a Scargle period analysis of all the data from JD\,533 to 538,
after subtracting the mean and linear trend from the light curves. The
power spectrum in Fig.\,\ref{f-power} has its highest peak at a period
of 0.07822(2)\,d, plus its one cycle/day aliases, which we
interpret as the signal from the superhumps. The superhump period
error estimate is derived using the method described by
\citet{schwarzenberg-czerny91-1}. A phase diagram of the data folded
on this period is shown in the top panel of Fig.\,\ref{f-folded} which
illustrates that both the superhump period as well as the morphology
of the superhump variation remain very stable throughout JD\,533 to
538.  While the values of $\Psh$ from the two methods are in close
agreement, it is not unusual for O--C analysis to give a more accurate
method of tracking periodic waves in dwarf novae as it is less
troubled by changes in amplitude and peak shape than period analysis
techniques \citep{pattersonetal02-2}. Thus in view of the consistency
of the O--C analysis, we take our value of $\Psh$ from that analysis.

Superhumps were still in evidence on JD\,542
(Fig.\,\ref{f-lightcurves}) and the measurement of a single time of
superhump maximum showed it to be broadly consistent with the above
superhump ephemeris. However by JD\,543 the superhumps were much less
coherent and there was considerable scatter in the data. Under these
circumstances it was not possible to measure superhump maximum
timings.

\citet{aungwerojwitetal06-1} gave two possible superhump periods,
derived from a short observing run during the first observed
superoutburst of HS\,0417, 0.0752\,d and 0.0772\,d, with a slight
preference for the longer period. Our value of $\Psh=0.07822(2)$\,d
determined from a much larger data set confirms their choice, and
supersedes their less accurate superhump period. Our value of $\Psh$
also agrees with the results independently obtained by
\citet{katoetal09-1} and \citet{katoetal10-1} for the 2008 and 2010
superoutbursts.

\begin{table}
\caption[]{Recorded outbursts of HS\,0417. Given are the UT date and
  Julian date of the outburst, the time interval between two
  consecutive outbursts, the peak magnitude, and the outburst
  duration. \label{t-outbursts}.}
\setlength{\tabcolsep}{0.95ex}
\begin{flushleft}
\begin{tabular}{ccccc}
\hline\noalign{\smallskip}
Date (UT) & JD--2450000 & Interval (d) & $m_\mathrm{max}$ & Duration
(d) \\
\hline\noalign{\smallskip}
2005 Apr 10 & 3470.6 &       & 13.6 & $\sim8$\\
2006 Feb 15 & 3782.4 & 311.8 & 15.6 & $<8$\\
2006 Sep 12 & 3991.4 & 209.0 & 14.6 & $\sim7$\\
2007 Mar 7  & 4167.4 & 176.0 & 14.8 & $<7$\\
2007 Oct 17 & 4390.6 & 223.2 & 14.7 & $<7$\\
2008 Mar 3  & 4529.5 & 138.9 & 13.2 & $\sim16$\\
2008 Aug 14 & 4693.4 & 163.9 & 15.7 & $<6$\\
2009 Feb 25 & 4888.3 & 194.9 & 15.9 & $<8$ \\
2009 Jun 11 & 4993.7 & 105.4 & 14.5 & $<7$ \\
2009 Dec 18 & 5184.3 & 190.6 & 14.4 & $<7$ \\
2010 Sep 2  & 5441.6 & 257.3 & 13.5 & $\sim16$\\
\hline\noalign{\smallskip}
\end{tabular}
\end{flushleft}
\end{table}

\subsection{Orbital humps and the orbital period}
Close examination of the superhump profiles reveals secondary humps in
some cases and occasionally these appear as distinct secondary
peaks. This was particularly obvious during JD\,533 as indicated by
the arrows in Fig.\,\ref{f-lightcurves}. It was evident that the
secondary humps appeared to move with respect to the underlying
superhump, suggesting they had a slightly shorter period. One possible
explanation is that the secondary humps represent an orbital hump,
which if true would allow the orbital period to be determined. To
explore this possibility further we measured the times of maximum of
nine such secondary humps, according to the Kwee and van Woerden
method. As with the previous superhump period analysis, we again
assigned orbit numbers which best fitted the assumption of a constant
orbital period. We found that the maxima fitted well a constant
orbital period $\Porb=0.07531(8)$\,d (108.45\,min), with a orbital
hump ephemeris thus:

\begin{equation}
\mathrm{JD(max)} = 2454533.40296 + 0.07531(8) \times E
\end{equation}

\begin{figure}
\centerline{\includegraphics[angle=-90,width=0.6\columnwidth]{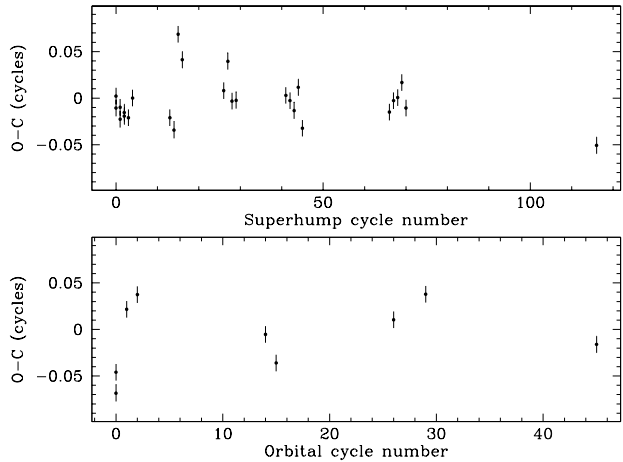}}
\caption{\label{f-oc} Top panel: Observed minus computed (O--C) times
  for the superhump maxima measured from the light curves in
  Fig.\,\ref{f-lightcurves}.  Bottom panel: O--C times of the orbital
  hump maxima.}
\end{figure}

The orbit number, the measured times of orbital hump maximum and the
O--C residuals relative to the above orbital hump maximum ephemeris
are shown in the bottom panel of Fig.\,\ref{f-oc}.

As an independent test on our measurement of $\Porb$ from the timing
of the orbital hump maxima, we pre-whitened the data from JD\,533 to
538 with $\Psh=0.07822$\,d, the (constant) superhump period prevailing
during this interval (Sect.\,\ref{s-psh}).  The Scargle periodogram
calculated from the pre-whitened data has two major peaks at
0.07581(51)\,d and 0.0819(58)\,d. These appear to be aliases as
further pre-whitening with either one of these signals causes both to
disappear from the power spectrum. We suggest that the shorter-period
signal is due to the orbital hump and hence represents the orbital
period. We reject the longer-period signal as being due to the orbital
hump as it would imply $\Porb>\Psh$, which would be very unusual for
an SU\,UMa type dwarf nova. The values of $\Porb$ derived from the two
methods are consistent with each other, given their errors, however we
prefer the value given by the linear orbital hump ephemeris for  the 
reasons given Sect.\,\ref{s-psh}.

\citet{aungwerojwitetal06-1} determined two possible orbital periods
of HS\,0417 from a number of photometric time series obtained in
quiescence, 0.730\,d and 0.0763\,d, with some preference for the
longer period. Their data suffered from severe fine-structured
aliases due to the sparse sampling of the photometry. Our orbital
period, $\Porb=0.07531(8)$\,d agrees broadly with the preferred value
of \citet{aungwerojwitetal06-1}. While we feel confident about our
orbital period determination based on the detection of orbital humps
superimposed on the superhumps, a spectroscopic confirmation of
$\Porb$ would be desirable. 

\begin{figure}
\centerline{\includegraphics[width=0.6\columnwidth]{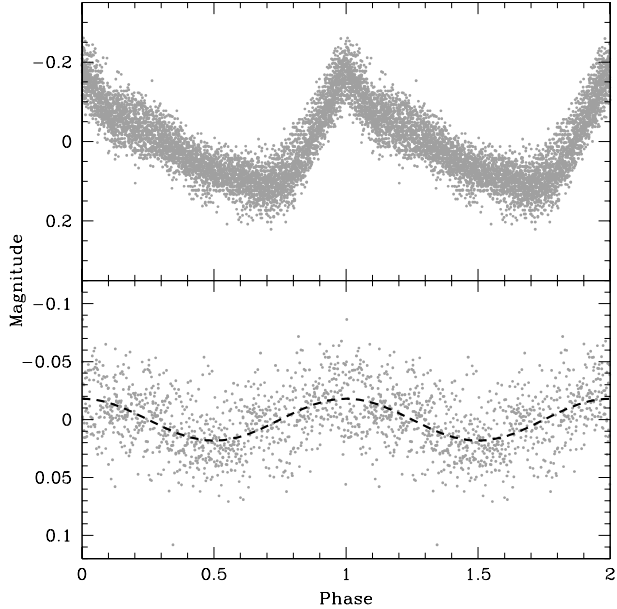}}
\caption{\label{f-folded} Top panel: all data from JD\,=\,533 to 538
  folded on the superhump period $\Psh=0.07822$\,d. Bottom panel: the
  data from JD\,=\,530 folded on $\Porb=0.07531$\,d. The dashed line
  shows a sine-fit to the folded data.}
\end{figure}

Having established a value of $\Porb$, we can now address the
low-amplitude modulations seen during the decline from the precursor
outburst on JD\,530 (Fig.\,\ref{f-outburst}, \ref{f-lightcurves}) and
ask whether they are associated with the orbital humps seen during the
subsequent superoutburst. Orbital humps have been seen during the
early stages of superoutbursts in a number of SU\,UMa systems
\citep{odonoghue00-1}, especially in the short orbital period dwarf
novae WZ\,Sge \citep{pattersonetal81-1}, AL\,Com \citep{katoetal96-1}
and HV\,Vir \citep{leibowitzetal94-1}, but also in normal SU\,UMa
systems, e.g. V1040\,Cen (also known as RX\,J1155.4-5641,
\citealt{pattersonetal03-1}). Since HS\,0417 was fading rapidly during
the photometry conducted on JD\,530, we removed this linear trend from
the data and then performed a Scargle period analysis on the data. The
resulting power spectrum has a rather broad main peak at
0.0733(69)\,d, which is consistent with our value of $\Porb$
determined above. A phase diagram with the data folded on this period
is shown in the bottom panel of Fig.\,\ref{f-folded}. We conclude that
it is likely that the low-amplitude humps seen during the decline from
the trigger outburst are orbital humps. 

\section{Discussion}
\label{s-discussion}

Taking the measured values $\Porb=0.07531(8)$\,d and $\Psh=0.07824(2)$
d allows the fractional superhump period excess $\epsilon=(\Psh-\Porb)
/ \Porb$ to be calculated as 0.037. This value is consistent with the
range of  $\epsilon$ observed in other SU\,UMa dwarf novae with
similar $\Porb$ \citep{pattersonetal05-3}. Measuring $\epsilon$
provides a way to estimate the mass ratio, $q=\Msec/\Mwd$, of a CV,
and following \citet{pattersonetal05-3} we find $q\simeq0.16$ for
HS\,0417. 

The appearance of very long-period superhumps (0.0856(88)\,d) during
the rise to the superoutburst maximum is unusual, as this phase is more
often characterised by the appearance of orbital humps
\citep{odonoghue00-1}. 

Our observations of this superoutburst of HS\,0417 clearly documented
the occurrence of a precursor outburst that was well-separated from
the main superoutburst. From the published superoutburst light curves,
it appears that precursor outbursts occur in some, but by no means
all SU\,UMa dwarf novae, and no global study on the presence or
absence of such events has been made so far. VW\,Hyi is probably the
SU\,UMa dwarf nova with the best-documented long-term light curve, and
precursor outbursts are seen during some, but not all superoutbursts
(e.g. \citealt{vogt83-2} and references therein), which could be
suggestive that the time lag between the precursor outburst and the
superoutburst is variable, leading sometimes to a smooth transition
from one to the other. Similarly, superoutbursts with and without
precursor outburst have been observed in TV\,Cor
\citep{uemuraetal05-1}.  A further complication is that whether the
precursor is seen clearly separated from the superoutburst can be a
function of the observed wavelength. \citet{pringleetal87-1} presented
a multi-wavelength study of VW\,Hyi where a distinct precursor
outburst was observed in the ultraviolet, and possibly in X-rays, but
not in the optical. Finally, precursor outbursts appear to occur in
other disc-accreting systems as well, such as e.g. the black-hole
binary XTE\,J1118+480 \citep{uemuraetal00-3, kuulkers00-1}.

Generally it is agreed that superhumps are the observational
manifestation of the precession of an eccentric accretion disc that
extends to the 3:1 resonance radius (\citealt{whitehurst88-1}, see
\citet{smithetal07-1} for a comprehensive 3D simulation). However, the
\textit{cause} of superoutbursts is still debated. In the
thermal-tidal instability model (TTIM, e.g. \citealt{osaki89-1}), the
accretion disc radius gradually grows throughout the superoutburst
cycle, until the disc radius reaches the 3:1 resonance radius during a
normal outburst, which triggers enhanced tidal dissipation in the disc
and leads the system into the superoutburst. Hence, in the TTIM, the
disc growth to the 3:1 radius triggers the superoutburst. In the
competing enhanced mass transfer model (ETMT,
e.g. \citealt{vogt83-2}), a feedback mechanism exists between the
accretion luminosity and the mass loss rate of the companion star. In
the ETMT, increased mass loss triggered by a normal outburst results
in the disc growing to the 3:1 radius, hence the onset of a
superoutburst is a consequence of the feedback
mechanism. The presence or absence of precursor outbursts holds
significant diagnostic potential to differentiate between the two
models. \citet{schreiberetal04-1} presented a series of models for
both the TTIM and ETMT, and concluded that the variety of
superoutbursts with and without precursor are better described by the
ETMT model. 

Substantially more information can be drawn from time-resolved
photometry covering the very early stages of a superoutburst, but
unfortunately very few such studies exist due to the time-critical
nature of securing the necessary observations. \citet{kato97-1}
analysed observations obtained during the precursor outburst of the
1993 T\,Leo superoutburst, and concluded that superhumps were already
present during the decline from the precursor. In contrast to this, no
significant modulations were seen during the precursor outbursts of
the 2003 superoutburst in GO\,Com \citep{imadaetal05-1} and during the
precursor outburst of the 2004 superoutburst of TV\,Cor
\citep{uemuraetal05-1}. Our observations of HS\,0417 clearly rule out
the presence of superhumps during the decline of its 2008
superoutburst, and suggest that instead orbital humps were present
during this phase. Superhumps with an unusually large fractional
period excess ($\epsilon=0.137$) developed subsequently very early
during the rise to the superoutburst maximum. The development of
superhumps during the 2008 precursor outburst and subsequent
superoutburst of HS\,0417 is consistent with the predictions of the
TTIM, where a normal outburst results in the growth of the accretion
disc to the 3:1 resonance radius, triggering the superoutburst.

\section{Conclusions}
We have obtained extensive photometry of the second recorded
superoutburst of the SU\,UMa dwarf nova HS\,0417. The superoutburst
was preceeded by a precursor outburst. Time-series analysis of our
data provides an accurate measurement of the superhump period,
$\Psh=0.07824(2)$\,d, and a robust estimate of the orbital period,
$\Porb=0.07531(8)$\,d. Combining both values gives a superhump period
excess of $\epsilon=0.037$, typical for SU\,UMa type dwarf novae in
this orbital period range, and subsequently an estimated mass ratio
$q=\Mwd/\Msec\simeq0.16$.

Our observations add to the small number of precursor outbursts that
are well-documented in terms of time-resolved photometry. Our data
rules out that superhumps were present during the decline from
precursor outburst, and strongly suggests that the light curve of
HS\,0417 was instead modulated on the orbital period during that
phase. Superhumps developed during the rise to the main
superoutburst, but initially exhibited an unusually large period excess
of $\epsilon=0.137$ ($\Psh=0.0856(88)$\,d), but then settled into a
regime of constant $\Psh$ for at least six days. 

Our observations demonstrated that the rapid response by a network of
small telescopes can provide detailed insight into the still poorly
known dynamics of the temporal evolution superoutbursts in compact
binaries with extreme mass ratios, and we encourage a more systematic
study of these events.

\bibliographystyle{elsarticle-harv.bst}
\bibliography{aamnem99,aabib,proceedings,submitted}







\end{document}